# Revised Orbital Parameters for Planets

André LE FLOCH

University of Tours, Department of Physics  37200 Tours, France

e-mail:    lefloch@univ-tours.fr

**Abstract.**   Orbital parameters of planets are fitted directly to an appropriate set of observations. It is shown how to use the rigorous Deming method combined with a numerical integration of gravitation equations. In all, 65 parameters of the nine planets (masses and initial positions and velocities at J2000) are listed. The complete set of their standard errors, and the associated variance-covariance matrix are presented for the first time. Derived parameters as the Solar gravitational quantity $G \times Mo$, the Astronomical distance $AU$, and the light time $\tau_A$ (for one $AU$) are re-evaluated. It is demonstrated that a direct fit using *time units* overcomes the very high correlation (99.99996 %) between the gravitational constant $G$ and the Solar mass $Mo$. Much more accurate values for these fundamental quantities are obtained:

| parameter | value | units | standard-deviation |
|---|---|---|---|
| $G \times M_o$ | $1.327731601427 \times 10^{+20}$ | m³/s² | $2.2 \times 10^{+09}$ |
| $AU$ | $1.496206824595 \times 10^{+11}$ | m | 2.4 |
| $\tau_A$ | 499.0808756753 | s | $8.1 \times 10^{-09}$ |

# 1. Introduction

A great debate has been started about the future climate of the Earth. In this context, astronomy should provide long-term predictions for the Earth orbit (for instance values of the perihelion and aphelion).

To undertake these extrapolations, we need first a complete set of orbital parameters for planets (including all masses and initial positions and velocities) in order to perform a numerical integration of gravitation equations. Also, we need the variance-covariance matrix associated with these orbital parameters, because we must calculate the statistical errors propagated on these predictions from the variance-covariance matrix.

Unfortunately, these starting data are not all available in regular astronomical literature. Of course, recent astronomical compilations of Yoder (1995) and Simon *et al* (1998) provide good values for masses of planets. However, initial positions and velocities of planets (at J2000) are not published in this literature. Moreover, the usual standard errors connected with these parameters, are missing in these compilations, as well as their variance-covariance matrix. So, the aim of this paper is to provide a complete and self-consistent set of orbital parameters for planets, including all standard errors, and also the associated variance-covariance matrix.

As pointed out by many authors belonging to different disciplines (Wentworth 1965a,b, Zare *et al*. 1973), the reduction of observed quantities should satisfy the two following requirements: (*i*) The fitting procedure must be a direct least-squares approach yielding the minimum-variance unbiased estimates; (*ii*) The method should maintain the physical or mechanical meaning of parameters.

In the case of measurements of planet positions, we are dealing with *three* variables (time, right ascension, and declination), and with their *three* corresponding uncertainties. So, fitting methods similar to the one of Oesterwinter and Cohen (1972), where errors with time are ignored, are not very rigorous (time is not an exact variable as a quantum number). Also, in view of the second requirement (*ii*), fitting methods using time polynomial expansions are not appropriate for a good numerical treatment of these astronomical data. First, their higher order coefficients have no physical meaning. In addition, these polynomial expressions cannot be used for extrapolations, because the relevant quantities are becoming infinite with time. So, we believe that the rigorous fitting procedure of Deming (1964, see also Wentworth's papers (1965a,b)), combined with a numerical integration of gravitation equations satisfies these two requirements.

Thus, our plan of campaign is quite clear: (*1*) First, we shall begin by performing a coherent data-file of observations for the Sun and all planets. (*2*) Then, carry out the fitting procedure connected with these data. (*3*) Finally, extend astronomical data from these results.

## 2. Description of data

An input data-file containing 3549 lines was carried out as indicated in Table *I*: First, Saturn + Uranus + Neptune + Pluto observations performed at the La Palma Observatory between 1984 and 1993 were extracted from the *Carlsberg Meridian Catalogue*. In addition, 61 observations of Saturn obtained between 1970 and 1978 at the Paris and San Fernando Astrolabes (Chollet *et al,* 1973, Debarbat *et al,* 1975, Sanchez *et al,* 1992) were also included; as well as 26 observations of Pluto recorded at the Moscow and Mt Maidanak Observatories (Dolganova *et al,* 1993).

In order to fit simultaneously all the orbital elements of planets, these observations were extended between year 1930 (year 1900 for Pluto) and year 2000, by selecting periodic data from the "*Connaissance des Temps*" ephemeris for all planets, and also for the Sun (Berthier *et al.,* 1999). Table *I* shows the interval (in days) between successive data for each object.

Thus, we get in all a consistent data set of 3549 observations, allowing to fit correctly the orbital parameters for planets.

# 3. Calculation of planet positions

Cartesian coordinates of planet positions are first computed in the standard "J2000.0" heliocentric system. For all the relevant time values, these quantities have been determined from a numerical integration of the usual following gravitation equations:

$$\ddot{\mathbf{r}}_j + \frac{G}{\gamma_j}.(M_o + M_j)\frac{\mathbf{r}_j}{|r_j|^3} = \sum_{k \neq j} G.M_k (\frac{\mathbf{r}_k - \mathbf{r}_j}{|r_{jk}|^3} - \frac{\mathbf{r}_k}{|r_k|^3})$$

*Equation (3.1)*

Where $\mathbf{r}_j$ represents the heliocentric position of the planet $j$ having a mass $M_j$, $G$ is the gravitational constant, and $M_o$ the Solar mass at the standard time J2000.0. We put $r_j = |\mathbf{r}_j|$, and so on.

Remember that refined effects such as the advance of the perihelion of Mercury could be explained by General Relativity corrections. As discussed in the Moller's text-book (Moller, 1972, p. 493), it is quite simple to introduce the following $\gamma_j$ term in eqs. (3.1):

$$\gamma_j = \left(1 - \frac{\left(\frac{2GM_0}{r_j} - (u_j)^2\right)}{c^2}\right)^{1/2} \qquad (3.2)$$

where $u_j$ designates the heliocentric velocity of the planet, and $c$ the usual light speed.

In our case (with nine planets), equations (3.1) yield 54 first order differential equations. These coupled equations have been integrated, numerically, from the $10^{th}$ degree Fehlberg's algorithm (Fehlberg, 1969) with a step of integration equal to h=3 ×$10^4$ sec ($\cong$ 0.347 day).

Then, heliocentric coordinates of the center of the Earth are extracted from positions of the barycentric Earth-Moon system, by using Lunar data of Chapront-Touzé *et al* (1988).

The apparent geocentric direction of the planet (or the Sun) at time $t_i$ is then obtained from the geometric positions of the Earth and the object at the retarded time $t_i - \tau_i$ (backward integration), where $\tau_i$ is the time of flight of the photon (Danjon, 1959).

Finally, the apparent angular coordinates $\alpha_i$ (right ascension) and $\delta_i$ (declination) are calculated, and corrected for precession from data of Williams (1994), and for nutation via the routine of Kinoshita *et al* (1990).

# 4. The fitting procedure

The orbital parameters of planets have been simultaneously fitted to the set of data described in section 2 by using the fitting procedure of Deming (1964), applied to physical sciences by Wentworth (1965a,b).

Note that this method allows to minimize the sum of the weighted squares of residuals:

$$S = \sum_{i=1}^{n} \left[ Wt_i (t_i - \overline{t_i})^2 + W\alpha_i (\alpha_i - \overline{\alpha_i})^2 + W\delta_i (\delta_i - \overline{\delta i})^2 \right]$$

<div align="center">equation (4.1a)</div>

where $t_i, \alpha_i, \delta_i$ and $\overline{t_i}, \overline{\alpha_i}, \overline{\delta_i}$ represent the observed and calculated quantities, respectively. The statistical weights are given by:

$$Wt_i = \frac{\sigma_o^2}{(\sigma(t_i))^2} \; ; \quad W\alpha_i = \frac{\sigma_o^2}{(\sigma(\alpha_i))^2} \; ; \quad W\delta_i = \frac{\sigma_o^2}{(\sigma(\delta_i))^2}$$

<div align="center">equation (4.1b)</div>

where $\sigma(t_i), \sigma(\alpha_i), \sigma(\delta_i)$ are the corresponding uncertainties, and $\sigma_o^2$ the arbitrary variance of unit weight.

Note that the goodness of the fit is given by the reduced variance

$$\sigma = \left(S/(n-p)\right)^{1/2} \qquad (4.2)$$

where $n$ is the total number of observations, and $p$ the number of fitted parameters. In our case of three variables, it is convenient to put $\sigma_o^2 = 1/3$, because a perfect fit corresponds to $\sigma \cong 1$. Of fundamental importance is the variance-covariance matrix

$$V = \sigma^2 N^{-1}$$

where $N$ is the normal equation matrix whose elements are given in *appendix A*. Remember that diagonal elements of $V$ represent the squares of standard errors of the fitted parameters. $V$ is also essential for calculations of the errors propagated to astronomical quantities.

## 5. Results

Our final fit yields a reduced variance $\sigma = 0.6$. This rather good value indicates that all modulus of residuals are nearly smaller than their corresponding uncertainties. We note that it is the case of those of Mercury, suggesting that, as expected, the General Relativity correction $\gamma_j$ (eq. 3.2) is convenient. In addition, 65 orbital parameters have been simultaneously fitted to the preceding 3549 input data. They are collected, for each planet, along with their statistical errors, in Tables *II-III*.

Because of the very high correlation coefficient (99.99996 %) between the gravitational constant $G$ and the Solar mass $Mo$, time units (*TU*), with $G = 1$ and $c=1$, are the most appropriate ones for these calculations (Synge, 1966). These results appear on the left part of Tables *II-III*. Physical units (*PU*) have been also used, allowing to fit the gravitational constant *G*. The corresponding quantities are displayed on right part of these tables. Some recent data of Yoder (1995) and of Simon *et al* (1998) are also reported for comparison. We emphasize that all these fitted orbital parameters (positions and velocities) are those for the standard epoch J2000.0: January 1$^{st}$ 2000 at 12h (Julian day 2 451 545.0).

Table *II* indicates that the solar mass *Mo* is determined with a very good accuracy of about $2\times10^{-11}$ by using time units, whereas this accuracy is only $2\times10^{-4}$ with physical units, because of the high correlation coefficient
between *G* and *Mo*. Hence, the accuracy of the gravitational constant is also of about $2\times10^{-4}$.

We note that our determinations of *G* and *Mo* (with physical units) are in agreement with previous data of Yoder and Simon *et al*. Of fundamental importance is also the product *G*×*Mo*, which appears at the bottom of Table *II*. Note that our value of *G*×*Mo* has the same accuracy ($2\times10^{-11}$) as our *TU* value of *Mo*, because masses in *PU* and in *TU* are connected by relation (Synge, 1966):

$$M(PU) = M(TU) \times \frac{c^3}{G} \qquad (5.1)$$

So, *G*×*Mo* in *PU* is merely the product of the cube of the speed of light (which is an exact value) by *Mo* in *TU*. In other words, this important parameter *G*×*Mo* is fitted directly to the data, if *TU* are used.

Table *II* shows that there is a strong disagreement between our value and the one reported by Yoder (the relative difference is of about $1.5\times10^{-4}$). First, we remark that their value was not fitted directly to their data, but calculated via the Kepler's third law *G*×*Mo* = $k^2(AU)^3 d^{-2}$. So, as the Gaussian constant *k* and the Julian day *d* are exact quantities, we believe that it is their *AU* value which is not as accurate as they wrote. This is due to the fact that parameters involving both *G* and *Mo* could *not* be fitted simultaneously with a very good accuracy, because of the very important correlation coefficient between *G* and *Mo*. In other words, previous data for *G×Mo*, *AU*, and $\tau_A$ (light time for one *AU*) reported in literature contain *ineluctably* this "*correlation*" error of about $2\times10^{-4}$ (that we have obtained with *PU*). So, the most accurate way for obtaining these fundamental quantities is: (1) Fit the data by using *TU*; (2) calculate *G×Mo* (with eq. 5.1); (3) compute *AU*, and then $\tau_A$ from the Kepler's third law. Our revised values for these parameters appear in Table *II*. Their precision is also of about $2\times10^{-11}$ (as for *Mo* with *TU*).

Note also that we have carried out an attempt to detect the time-dependence of the Solar mass *Ms* via the relation (at time $t_i$ ) *Ms=Mo× exp(-ρ×$t_i$)*. As expected (because our data-file contains only observations for one century), the determination of this new parameter $\rho$ was not statistically significant. Thus, this coefficient was fixed to zero in all our fits.

From Table *III*, consider now masses of planets. For bigger ones, their masses are determined with an accuracy of about 0.02 % for Jupiter and Saturn, 0.7 % for Uranus, and 1 % for Neptune. For small planets, the precision is 0.2 % for masses of Venus and for the Earth-Moon system, 0.6 % for Mars, and only 4% for Mercury. In the case of Pluto, our accuracy was so bad (a standard error as great as the parameter) that this mass was fixed in our fits to the value reported by Simon *et al*. (1998).

Determinations of masses are very difficult because derivatives $\dfrac{\partial \alpha}{\partial M}$ and $\dfrac{\partial \delta}{\partial M}$ are very small. Note also that masses in *PU* and *TU* are connected by relation (5.1). Except for Pluto (which has a fixed value), this relationship is satisfied for all planets, within the combined standard errors.

From results collected in Table *III*, we have also calculated in Table *IV* the accuracy $\dfrac{\Delta r_o}{r_o}$ and $\dfrac{\Delta u_o}{u_o}$ of the initial values of positions and velocities of planets. We note that these precisions are very good: about $7\times10^{-8}$ for the barycentre of the Earth-Moon system in the best case, few units of $10^{-7}$ for Jupiter, Mars and Venus, and of the order of $10^{-6}$ for the other planets, even for Pluto. In other words, orbits of planets depend strongly of the initial positions and velocities, but only a little of their masses. The case of Mercury and Pluto illustrates this remark: they have nearly similar masses, but very different trajectories.

Finally, our results are compared (in Table *III* ) with those previously obtained by Yoder (1995) and Simon *et al* (1998) for masses, even if errors are rarely reported by these authors. Nevertheless, there is a rather good agreement between all these masses.

# 6. Conclusion

This paper provides a self-consistent set of orbital parameters, including their standard errors, which have been simultaneously fitted to appropriate data. By combining these parameters with the relevant variance-covariance matrix, it will be possible to perform long-term predictions of astronomical data, along with their propagated errors.
Concerning parameters, we have shown that it is possible to overcome the very high correlation between the gravitational constant *G* and the Solar mass *Mo*. So, we have proposed values for the fundamental quantities *G×Mo*, *AU*, and $\tau_A$, which are much more accurate than those previously reported in literature.

In a near future we plan, as far as possible, to add new observations in our data-file (for instance those of Mars and Jupiter recorded at the San Fernando Observatory, when these raw data will be available). However, we believe that these forthcoming extended fits will bring only minor variations to parameters listed in the present work.
(preceding data-files are available on request from the author).

# Appendix A: Elements of the normal equation matrix

This section collects, without demonstrations, exact relationships allowing to code computing programs for fitting orbital parameters for planets. It follows the original work of Deming (1964), revisited by Wentworth (1965 a,b). The case of observations of positions of planets corresponds to the method detailed in Deming's book (p. 50), for three variables (*t*, $\alpha$, $\delta$) connected by two "*condition equations*" for each object, and with many parameters. Essentially, the method of Lagrange multipliers allows to minimize the sum of the weighted squares of residuals (see eq. 4.1a).

Recall that it is an iterative method. Thus, the correction-vector $\Delta A$ (with components $\Delta a_j$) to the initial approximate parameter vector $A^o$ (with components $a_j^o$) is given by:

$$\Delta A = N^{-1} \Delta W$$

So, corrected parameters are (*p* is the total number of parameters):

$$a_j = a_j^o - \Delta a_j \qquad (j = 1, 2, 3, ...., p)$$

Elements of the normal equation matrix $N$ (square-symmetric of dimensions *p*×*p*) are given by the following equations:

$$N(j,k) = \sum_i \left( \frac{\partial \alpha_i}{\partial a_k} \times u_i(j) + \frac{\partial \delta_i}{\partial a_k} \times v_i(j) \right)$$

Remark that dimensions of the normal equation matrix $N$ (*p*×*p*) are quite small. We do not need to handle "enormous" matrix as reported in calculations of Oesterwinter and Cohen (1972).

The *p* elements of the error-vector $\Delta W$ are:

$$\Delta W(j) = \sum_i \left( V\alpha_i^o \cdot u_i(j) + V\delta_i^o \cdot v_i(j) \right)$$

where $V\alpha_i^o$ and $V\delta_i^o$ represent the initial residuals of right ascension and declination. Summations are extended over all the *n* observations.

Furthermore, the $u_i(j)$ and $v_i(j)$ terms are computed as follows:

$$u_i(j) = s_i \left( q_i \frac{\partial \alpha_i}{\partial a_j} - r_i \frac{\partial \delta_i}{\partial a_j} \right)$$

$$v_i(j) = s_i \left( p_i \frac{\partial \delta_i}{\partial a_j} - r_i \frac{\partial \alpha_i}{\partial a_j} \right)$$

with (see also eq. 4.1b)

$$p_i = (1/W\alpha_i) + \left( \frac{\partial \alpha_i}{\partial t_i} \right)^2 / Wt_i$$

$$q_i = (1/W\delta_i) + \left( \frac{\partial \delta_i}{\partial t_i} \right)^2 / Wt_i$$

$$r_i = \left( \frac{\partial \alpha_i}{\partial t_i} \frac{\partial \delta_i}{\partial t_i} \right) / Wt_i$$

$$s_i = \left( p_i q_i - r_i^2 \right)^{-1}$$

Finally, residuals of the three variables ($Vt_i, V\alpha_i, V\delta_i$) could be calculated via the Lagrange multipliers $\lambda_i$ and $\mu_i$:

$$V\alpha_i = \lambda_i / W\alpha_i \; ; \; V\delta_i = \mu_i / W\delta_i \; ; \; Vt_i = (\lambda_i + \mu_i)/Wt_i$$

with

$$\lambda_i = s_i(d_i r_i - b_i q_i)$$

$$\mu_i = s_i(b_i r_i - d_i q_i)$$

$$b_i = V\alpha_i^o - \sum_{j=1}^{p} \frac{\partial \alpha_i}{\partial a_j} \Delta a_j$$

$$d_i = V\delta_i^o - \sum_{j=1}^{p} \frac{\partial \delta_i}{\partial a_j} \Delta a_j$$

Then, we can calculate the reduced variance σ and the variance-covariance matrix $V$. All preceding relationships have been used (and coded) in our work.
Remark that all derivatives appearing in preceding equations have been calculated numerically.

# Acknowledgements

We are very grateful to Dr. Kinoshita for supplying his personal nutation routine, and Olivier Boebion for his help during computing. We would like also to thank Dr. J. Llibre for his advice concerning the use of the Fehlberg's algorithm, Dr. S. Débarbat for providing her original data about Saturn, and Dr. M. Sanchez for his welcome at San Fernando Observatory.

# References

Berthier,J. and Descamps,P. : 1999, *Connaissance des Temps, Institut de*

*Mécanique Céleste et de Calcul d'Ephémérides*, Paris.

Carlsberg Meridian Catalogue: 1997, *Observations of positions of stars and planets*, La Palma, Nr. 9 (CD-rom).

Chapront-Touzé,M., and Chapront,J. : 1988, 'A semi-analytical lunar ephemeris for historical times', *Astron. Astrophys.* **190**, 342-352.

Chollet,F., Choplin,H., Débarbat,S., Feissel,M., Lam,S.K., : 1973, 'Observations de Saturne effectuées à l'astrolabe de Paris en 1971-1972', *A&A* **26**, 141-142.

Débarbat,S. : 1975, 'Observations de Saturne effectuées à l'astrolabe de Paris en 1972-1973', *A&A Suppl.* **19**, 389-393.

Danjon,A., : 1959, *Astronomie Génerale*, Sennac Editeur (Paris).

Deming,W.,E. : 1964, *Statistical Adjustment of Data*, Dover Publications, Inc. (New York).

Dolganova,E.V., Kuimov,K.V., Shokin,Y.A. : 1993, 'Positional observations of Pluto in 1969, 1970, 1989, 1990, 1991', *Astron. Lett.* **19**, 397-399.

Fehlberg,E. : 1969, 'Klassische Runge-Kutta-Formeln funfter und siebenter ordnung mit schrittweiten kontrolle', *Computing*, 4, 93-106.

Kinoshita,H., and Souchay,J. : 1990, 'The theory of the nutation for the rigid earth model at the second order', *Celest. Mech.* **48**, 187-265.

Moller,C. : 1972, *The Theory of Relativity*, Clarendon Press (Oxford)

Oesterwinter,C., and Cohen,C.J. : 1972, 'New orbital elements for Moon and Planets', *Celest. Mech.* 317-395.

Sanchez,M., and Débarbat,S. : 1992, 'Corrections aux ascensions droites et aux déclinaisons de Saturne entre 1970 et 1978', *A&A Suppl.* **95**, 371-377.

Simon,J.L., Chapront-Touzé,M., Morando,B., Thuillot,W. : 1998, *Introduction aux Ephémérides Astronomiques*, EDP Sciences (Paris).

Synge,J.L. : 1966, *Relativity: The General Theory*, North-Holland Publishing Company (Amsterdam).

Wentworth,W.E., : 1965a, 'Rigorous least squares adjustment *I*', *J. of Chem. Educ.* **42**, 96-103.

Wentworth,W.E., : 1965b, 'Rigorous least squares adjustment *II*', *J. of Chem. Educ.* **42**, 162-167.

Williams,J.G. : 1994, 'Contributions to the Earth's obliquity rate, precession, and nutation', *Astron. J.*, **108**, 711-724.

Yoder,C.F., : 1995, *Astrometric and Geodetic Properties of Earth and the Solar System,* American Geophysical Union (Pasadena)

Zare,R.N., Schmeltekopf,A.L., Harrop,W.J., Albritton, D.L., :1973, 'A direct approach for the reduction of diatomic spectra to molecular constants', *J. Mol. Spectrosc*. 46, 37-66.

Table I.  Description  of  data.

___________________________________________________________________

| object | number of | | period | interval[d] (days) | number of data |
|---|---|---|---|---|---|
| | observations | predictions | | | |
| Pluto | 220 a + 26 b | 122 | 1900-2000 | 300 | 368 |
| Neptune | 813 a | 77 | 1930-2000 | 300 | 890 |
| Uranus | 712 a | 78 | id. | 300 | 790 |
| Saturn | 147 a + 61 c | 158 | id. | 300 | 366 |
| Jupiter | | 183 | id. | 250 | 183 |
| Mars | | 200 | id. | 200 | 200 |
| Venus | | 231 | id. | 150 | 231 |
| Mercury | | 294 | id. | 100 | 294 |
| Sun | | 227 | id. | 200 | 227 |

___________________________________________________________________

total input data = 3549
___________________________________________________________________

a): data from the Carlsberg Meridian Catalogue. b):data of Dolganova *et al*.

c): data from Paris-Astrolabe. d): interval (in days) between predictions

   of *Connaissance des Temps* (only for data before 1980).

Table II. Gravitational constant, Solar mass, and Astronomical Quantities.

______________________________________________________________

| parameter | time units (Gsec) | | physical units (Gsec,Gm,GKg) | | ref. |
|---|---|---|---|---|---|
| | value | std. err. | value | std. err. | |
| G | 1 | | 6.6721168E-11 | 9.6E-15 | a |
| | | | 6.67259 E-11 | 8.4E-15 | b |
| Mo | 4.927744515681E-15 | 8.0E-26 | 1.9899702E+21 | 2.9E+17 | a |
| | 4.920 | | | | d |
| | | | 1.9891 E+21 | | b |
| | | | 1.9889 E+21 | | c |

______________________________________________________________

| $G*Mo$ (SI units: $m^3/s^2$) | | | 1.327731601427E+20 | 2.2E+09 | a |
|---|---|---|---|---|---|
| | | | 1.3271243994 E+20 | 5.0E+10 | b |
| | | | 1.32712440 E+20 | | c |
| AU (m) | | | 1.496206824595E+11 | 2.4E+00 | a |
| | | | 1.4959787066 E+11 | 5.0E+01 | b |
| | | | 1.4959787061 E+11 | | c |
| $\tau_A$ (s) | | | 499.0808756753 | 8.1E-09 | a |
| | | | 499.00478370 | | b |
| | | | 499.00478353 | | c |

______________________________________________________________
______________________________________________________________

Values for J2000.0 (JD 2451545.0)

Note: a) present work; b) from Yoder(1995); c) data of Simon *et al* (1998); d) value of Synge(1966).

Table III. Orbital Parameters for Planets

______________________________________________________________

Mercury
______________________________________________________________

| parameter | time units (Gsec) | | physical units (GKg,Gm,Gsec) | | |
|---|---|---|---|---|---|
| | value | std. err. | value | std. err. | |
| M   | 8.5633E-22      | 2.9E-23  | 3.4440E+14       | 1.2E+13 | a |
|     |                 |          | 3.302 E+14       |         | b |
|     |                 |          | 3.3018E+14       |         | c |
| X   | -6.49314738E-08 | 1.4E-13  | -1.94658046E+01  | 4.3E-05 | a |
| Y   | -2.23231262E-07 | 1.0E-13  | -6.69230937E+01  | 3.1E-05 | a |
| Z   | -1.22762797E-08 | 2.0E-13  | -3.68033898E+00  | 6.3E-05 | a |
| Ux  | 1.23420487E-04  | 8.8E-11  | 3.70005539E+04   | 2.7E-02 | a |
| Uy  | -3.72477728E-05 | 7.3E-11  | -1.11665193E+04  | 2.3E-02 | a |
| Uz  | -1.43710226E-05 | 1.1E-10  | -4.30831412E+03  | 3.4E-02 | a |

______________________________________________________________

Venus
______________________________________________________________

| M   | 1.20865E-20     | 2.0E-23  | 4.88325E+15      | 8.4E+12 | a |
|---|---|---|---|---|---|
|     |                 |          | 4.8685 E+15      |         | b |
|     |                 |          | 4.8685 E+15      |         | c |
| X   | -3.58490929E-07 | 4.5E-14  | -1.07472864E+02  | 1.4E-05 | a |
| Y   | -1.62967002E-08 | 6.1E-14  | -4.88562247E+00  | 1.9E-05 | a |
| Z   |  2.04694016E-08 | 6.5E-14  |  6.13657094E+00  | 2.0E-05 | a |
| Ux  | 4.61015017E-06  | 1.6E-11  | 1.38208667E+03   | 4.8E-03 | a |
| Uy  | -1.17233301E-04 | 1.5E-11  | -3.51456604E+04  | 4.7E-03 | a |
| Uz  | -1.86840660E-06 | 2.1E-11  | -5.60133527E+02  | 6.6E-03 | a |

______________________________________________________________

Barycentric Earth-Moon system.
______________________________________________________________

| M   | 1.49946E-20     | 2.3E-23  | 6.05494E+15      | 9.9E+12 | a |
|---|---|---|---|---|---|
|     |                 |          | 6.0471 E+15      |         | b |
|     |                 |          | 6.0471 E+15      |         | c |
| X   | -8.84165321E-08 | 1.7E-14  | -2.65065886E+01  | 5.4E-06 | a |
| Y   |  4.82720728E-07 | 3.1E-14  |  1.44716027E+02  | 9.7E-06 | a |
| Z   | -5.71869163E-13 | 5.1E-14  | -1.76554413E-04  | 1.6E-05 | a |
| Ux  | -9.93720161E-05 | 6.3E-12  | -2.97909808E+04  | 2.0E-03 | a |
| Uy  | -1.82759998E-05 | 4.4E-12  | -5.47900403E+03  | 1.4E-03 | a |

```
Uz  1.51609383E-10        9.3E-12        4.59788384E-02   2.9E-03 a
```
___

Table III (continued)

___

Mars

___

```
M   1.60070E-21           8.2E-24        6.47358E+14      3.5E+12 a
                                         6.419  E+14              b
                                         6.4185 E+14              c

X    6.94079815E-07       6.9E-14        2.08079888E+02   2.2E-05 a
Y   -6.69565254E-09       2.6E-13       -2.00731349E+00   8.1E-05 a
Z   -1.72020850E-08       1.0E-13       -5.15705557E+00   3.3E-05 a

Ux   3.87883193E-06       2.7E-11        1.16284516E+03   8.4E-03 a
Uy   8.77275965E-05       7.9E-12        2.63000700E+04   2.5E-03 a
Uz   1.74247538E-06       1.3E-11        5.22380685E+02   4.1E-03 a
```
___

Jupiter

___

```
M    4.708566E-18         7.5E-22        1.901473E+18     4.2E+14 a
                                         1.89919 E+18              b
                                         1.8990  E+18              c

X    1.99691012E-06       4.5E-13        5.98658565E+02   1.4E-04 a
Y    1.46658954E-06       4.0E-13        4.39672474E+02   1.2E-04 a
Z   -5.07979691E-08       5.9E-13       -1.52288476E+01   1.8E-04 a

Ux  -2.63884390E-05       5.5E-12       -7.91105472E+03   1.7E-03 a
Uy   3.72185118E-05       8.7E-12        1.11578284E+04   2.7E-03 a
Uz   4.36585417E-07       1.2E-11        1.30885005E+02   3.8E-03 a
```
___

Saturn

___

```
M    1.408626E-18         2.7E-22        5.688479E+17     1.4E+14 a
                                         5.6864  E+17              b
                                         5.6860  E+17              c

X    3.19733321E-06       7.1E-12        9.58536398E+02   2.2E-03 a
Y    3.27895665E-06       4.0E-12        9.83006415E+02   1.3E-03 a
Z   -1.84188989E-07       6.0E-13       -5.52184725E+01   1.9E-04 a

Ux  -2.47942585E-05       4.2E-11       -7.43313111E+03   1.3E-02 a
Uy   2.24720635E-05       2.0E-11        6.73695495E+03   6.3E-03 a
Uz   5.94615340E-07       6.5E-12        1.78261186E+02   2.0E-03 a
```
___

Table III (continued)

___________________________________________________________

## Uranus
___________________________________________________________

| | | | | | |
|---|---|---|---|---|---|
| M | 2.16327E-19 | 1.4E-21 | 8.73649E+16 | 6.0E+14 | a |
| | | | 8.6634 E+16 | | b |
| | | | 8.6840 E+16 | | c |
| X | 7.20266288E-06 | 6.8E-12 | 2.15930420E+03 | 2.1E-03 | a |
| Y | -6.85450368E-06 | 1.8E-11 | -2.05492907E+03 | 5.5E-03 | a |
| Z | -1.18848871E-07 | 1.5E-12 | -3.56299889E+01 | 4.7E-04 | a |
| Ux | 1.54697452E-05 | 3.6E-11 | 4.63771401E+03 | 1.1E-02 | a |
| Uy | 1.54386933E-05 | 7.2E-11 | 4.62840099E+03 | 2.2E-02 | a |
| Uz | -1.43079191E-07 | 3.6E-12 | -4.28940385E+01 | 1.1E-03 | a |

___________________________________________________________

## Neptune
___________________________________________________________

| | | | | | |
|---|---|---|---|---|---|
| M | 2.51337E-19 | 2.6E-21 | 1.01637E+17 | 1.1E+15 | a |
| | | | 1.0280 E+17 | | b |
| | | | 1.0246 E+17 | | c |
| X | 8.39056779E-06 | 4.0E-12 | 2.51542922E+03 | 1.2E-03 | a |
| Y | -1.24728664E-05 | 1.0E-11 | -3.73927197E+03 | 3.1E-03 | a |
| Z | 6.34943840E-08 | 2.1E-12 | 1.90351466E+01 | 6.7E-04 | a |
| Ux | 1.48988775E-05 | 1.6E-11 | 4.46657177E+03 | 4.9E-03 | a |
| Uy | 1.02641971E-05 | 3.9E-11 | 3.07712750E+03 | 1.2E-02 | a |
| Uz | -5.54000997E-07 | 6.0E-12 | -1.66085313E+02 | 1.9E-03 | a |

___________________________________________________________

## Pluto
___________________________________________________________

| | | | | | |
|---|---|---|---|---|---|
| M | 3.5460E-23 | | | 1.4320E+13 | c |
| X | -4.92857601E-06 | 1.4E-11 | -1.47754991E+03 | 4.3E-03 | a |
| Y | -1.39536893E-05 | 1.9E-11 | -4.18321080E+03 | 5.8E-03 | a |
| Z | 2.91983922E-06 | 6.5E-12 | 8.75345780E+02 | 2.0E-03 | a |
| Ux | 1.74952860E-05 | 2.3E-11 | 5.24495462E+03 | 7.3E-03 | a |
| Uy | -8.88278738E-06 | 3.4E-11 | -2.66299265E+03 | 1.1E-02 | a |
| Uz | -4.11396503E-06 | 1.3E-11 | -1.23333565E+03 | 4.1E-03 | a |

___________________________________________________________
___________________________________________________________

(Values for J2000.0 (JD 2451545.0))

a) present work; b) from Yoder(1995); c) data of Simon *et al*(1998).

(extra significant figures are retained to avoid round-off errors)

Table IV. Errors for initial positions and velocities.

| planet | error for initial | |
|---|---|---|
| | position | velocity |
| Mercury | 6.2E-07 | 9.0E-07 |
| Venus | 1.4E-07 | 1.4E-07 |
| Earth-Moon | 6.8E-08 | 6.9E-08 |
| Mars | 1.1E-07 | 1.1E-07 |
| Jupiter | 2.5E-07 | 2.3E-07 |
| Saturn | 1.7E-06 | 1.3E-06 |
| Uranus | 1.7E-06 | 3.5E-06 |
| Neptune | 7.0E-07 | 1.9E-06 |
| Pluto | 1.6E-06 | 1.9E-06 |

Values for J2000.0 (JD 2451545.0)